\documentclass[12pt,preprint]{aastex}

\newcommand{\arcsecs}{\mbox{$^{\prime\prime}$}}

\newcommand{\hours}{\mbox{$^{h}$}}
\newcommand{\mins}{\mbox{$^{m}$}}

\shorttitle{GRB~001025A}
\shortauthors{Pedersen et al.}

\begin{document}

\title{Multi-Wavelength Studies of the Optically Dark Gamma-Ray Burst
001025A}

\author{
K.~Pedersen,\altaffilmark{1} K.~Hurley,\altaffilmark{2} J.~Hjorth,\altaffilmark{1} D.~A.~Smith,\altaffilmark{3} M.~I.~Andersen,\altaffilmark{4} L.~Christensen,\altaffilmark{4} T.~Cline,\altaffilmark{5} J.~P.~U.~Fynbo,\altaffilmark{1} J.~Goldsten,\altaffilmark{6} S.~Golenetskii,\altaffilmark{7} J.~Gorosabel,\altaffilmark{8} P.~Jakobsson,\altaffilmark{1} B.~L.~Jensen,\altaffilmark{1} B.~Milvang-Jensen,\altaffilmark{9} T.~McClanahan,\altaffilmark{5} P.~M\o ller,\altaffilmark{10} V.~Palshin,\altaffilmark{7} N.~Schartel,\altaffilmark{11} J.~Trombka,\altaffilmark{5} M.~Ulanov,\altaffilmark{7} D.~Watson\altaffilmark{1}
}

\altaffiltext{1}{Dark Cosmology Centre, Niels Bohr Institute, University of Copenhagen, Juliane Maries Vej 30, DK-2100 Copenhagen \O{}, Denmark; kp, jens, jfynbo, pallja, brian\_j, darach @astro.ku.dk}

\altaffiltext{2}{Space Sciences Laboratory, 7 Gauss Way, University of California, Berkeley, CA 94720-7450, USA; khurley@ssl.berkeley.edu}

\altaffiltext{3}{NSF Postdoctoral Fellow, University of Michigan, Department of Physics Ann Arbor, MI 48109, USA; donaldas@umich.edu}

\altaffiltext{4}{Astrophysikalisches Institut Potsdam, An der Sternwarte 16, 14482 Potsdam, Germany; mandersen, lchristensen @aip.de}

\altaffiltext{5}{NASA Goddard Space Flight Center
Code 661 Greenbelt, MD 20771, USA; Thomas.L.Cline, Timothy.P.McClanahan, Jacob.I.Trombka.1 @nasa.gov}

\altaffiltext{6}{The Johns Hopkins University Applied Physics Laboratory Laurel, MD 20723, USA; john.goldsten@jhuapl.edu}

\altaffiltext{7}{Ioffe Physico-Technical Institute, St. Petersburg, 194021, Russia; golen, val, mulanov @mail.ioffe.ru}

\altaffiltext{8}{IAA-CSIC, P.O. Box 03004, 18080 Granada, Spain and Space Telescope Science Institute, 3700 San Martin Drive, Baltimore, MD21218, USA;
jgu@laeff.esa.es}

\altaffiltext{9}{Max-Planck-Institut f{\"u}r extraterrestrische Physik, Giessenbachstr., D-85748 Garching; milvang@mpe.mpg.de}

\altaffiltext{10}{European Southern Observatory, Karl Schwarzschild-Strasse 2, D-85748 Garching bei M\"{u}nchen, Germany; pmoller@eso.org}

\altaffiltext{11}{{\it XMM-Newton} Science Operations Centre, ESA, Villafranca del Castillo, P.O. Box 50727, 28080 Madrid, Spain; nscharte@xmm.vilspa.esa.es}

\newpage

\begin{abstract}
We identify the fading X-ray afterglow of GRB~001025A from
{\it XMM-Newton} observations obtained 1.9--2.3~days, 2~years, and 2.5~years after
the burst. The non-detection of an optical counterpart to an upper limit of 
$R=25.5$, 1.20~days after the burst, makes GRB~001025A a ``dark'' burst.
Based on the X-ray afterglow spectral properties of GRB~001025A, we argue 
that some bursts appear optically dark because their
afterglow is faint and their cooling frequency is close
to the X-ray band. This interpretation is applicable to several of the few
other dark bursts where the X-ray spectral index has been measured.
The X-ray afterglow flux of GRB~001025A is an order of magnitude lower 
than for typical long-duration gamma-ray bursts. 
The spectrum of the X-ray afterglow can be fitted
with an absorbed synchrotron emission model, an absorbed thermal plasma model,
or a combination thereof.
For the latter, an extrapolation to optical wavelengths can be reconciled
with the $R$-band upper limit on the afterglow, without invoking any optical
circumburst absorption, provided the cooling frequency is close to
the X-ray band. Alternatively, if the X-ray afterglow is due to synchrotron
emission only, seven magnitudes of extinction in the observed $R$-band is
required to meet the $R$-band upper limit, making GRB~001025A
much more obscured than bursts with detected optical afterglows.
Based on the column density of X-ray absorbing circumburst matter, an SMC
gas-to-dust ratio is insufficient to produce this amount of extinction.
The X-ray tail of the prompt emission enters a steep
temporal decay excluding that the tail of the prompt emission is the onset of 
the afterglow.
To within the astrometric uncertainty, this afterglow was
coincident with an extended object, seen in a deep VLT $R$-band image,
which we identify as the likely host galaxy of GRB~001025A.
\end{abstract}


\section{INTRODUCTION}

One of the enduring mysteries of cosmic gamma-ray bursts (GRBs) is that of the
so-called ``dark'' bursts - those with no detected optical or near-infrared
afterglows. In the fireball model of GRBs \citep[for a review, see][]{mesz}, 
an explosive event produces a relativistically expanding blast wave.
Internal shocks within the blast wave are the source of the high
energy prompt emission while afterglows arise from external shocks when 
the blast wave run into an interstellar medium.
 
GRBs may be divided into ``long-duration'' and ``short-duration'' categories;
no unambigous detection of an optical afterglow has yet been made for any 
short-duration event \citep[e.g.][]{bloom05,hjorth05,gorosabel02,hur02a}
and only in one case has an X-ray afterglow been detected \citep{gehrels}. 
On the other hand, virtually all of the long-duration bursts display
fading soft X-ray afterglows \citep[e.g.][]{costa,depas}, and they have 
been shown to be associated with supernova explosions 
\citep{hjorth03b,stanek,malesani}.
 
For more than half of the long-duration GRBs no optical and/or 
radio afterglow has been detected \citep[e.g.][]{fyn01,lazatti02,berger};
however, rapid and systematic 
follow-up of bursts has decreased the relative fraction of dark bursts in recent
years \citep{rol,berger05}. Recently, \cite{pall} suggested using
the value of the optical-to-X-ray spectral index as a diagnostic for early 
identification of dark bursts, and they found five dark bursts (including 
GRB~001025A) in their sample of 52 bursts.

There are several possible explanations for dark bursts \citep{fyn01}. 
They could be at very high redshifts \citep{lr}, 
or they could be heavily obscured by dust in their 
host galaxies \citep[e.g.][]{groot98,taylor98,rp}.  
They could also occur in regions where the surrounding medium
is tenuous, far from the centers of their host galaxies, where the strong shocks 
required for particle acceleration could not form \citep{kp}. 

Observations of bursts detected by the {\it BeppoSAX},  
the {\it High Energy Transient Explorer}, and the {\it Rossi X-ray Timing Explorer}
spacecrafts have indeed shown that
bursts can be dark for a variety of reasons. Some bursts seem in fact to be dark due to 
extinction, e.g. GRB~000210 \citep{piro}, GRB~970828 \citep{djorg}, and
GRB~001109 \citep{josemaria}. Other bursts have faint optical afterglows 
either because of moderately high redshift, e.g., GRB~020124 
\citep{hjorth03a,berger}, or because they are intrinsically faint 
without extinction, e.g. GRB~980613 \citep{hjorth02} and GRB~021211 \citep{crew}. 
Analysis of bursts detected by {\it Swift} will undoubtly shed more
light on the nature of dark bursts. 
Already now it has become evident that afterglows of bursts 
detected by {\it Swift} on average
are fainter than afterglows from bursts detected by previous instruments 
(1.7~magnitudes in the 
optical/near-infrared (Berger et al., 2005) and a factor three in X-rays
Jakobsson, 2005). The rapid fading of some {\it Swift} bursts
demonstrates that at least these become intrinsically faint after the first few
hundred seconds \citep{tag}.

Here we examine the case of the dark burst GRB~001025A. 
We report on our temporal and spectral analysis of the
prompt emission, on our very deep optical upper limits on the afterglow,
and on the results of our temporal and spectral analysis of the X-ray afterglow.
Finally, we discuss the nature of GRB~001025A within the framework of the
fireball model, and the implications for dark bursts in general.

\section{OBSERVATIONS}

\subsection{GRB~001025A Prompt Emission}

GRB~001025A was first reported as an event detected 
by the \it Rossi X-Ray Transient
Explorer \rm (RXTE) All Sky Monitor (ASM) at 11405~s~UT October~25, 2000 
\citep{smith}.  Its
duration in the ASM 1.5--12 keV energy band was $\sim$15 s, and its peak flux
was $\sim$ 4 Crab ($\rm10^{-7}\, erg~cm^{-2}~s^{-1}$) in the 5--12 keV band.  
The event was detected by a single
camera and localized to a $\rm 4 \arcmin \times 1{\fdg}6$ error box.  
It was also observed by {\it Ulysses}, 
the X-Ray/Gamma-Ray Spectrometer 
(XGRS) aboard the \it Near Earth Asteroid Rendezvous \rm mission (NEAR),
and \it Wind \rm (Konus experiment) in the Interplanetary Network 
(IPN), and
triangulated to a preliminary $ \rm 5.6 \arcmin $ wide annulus which crossed 
the ASM error box to form a 25~arcminute$^2$ error box 
\citep{hur00a,hur00b}.

\begin{figure}
\epsscale{0.45}
\plotone{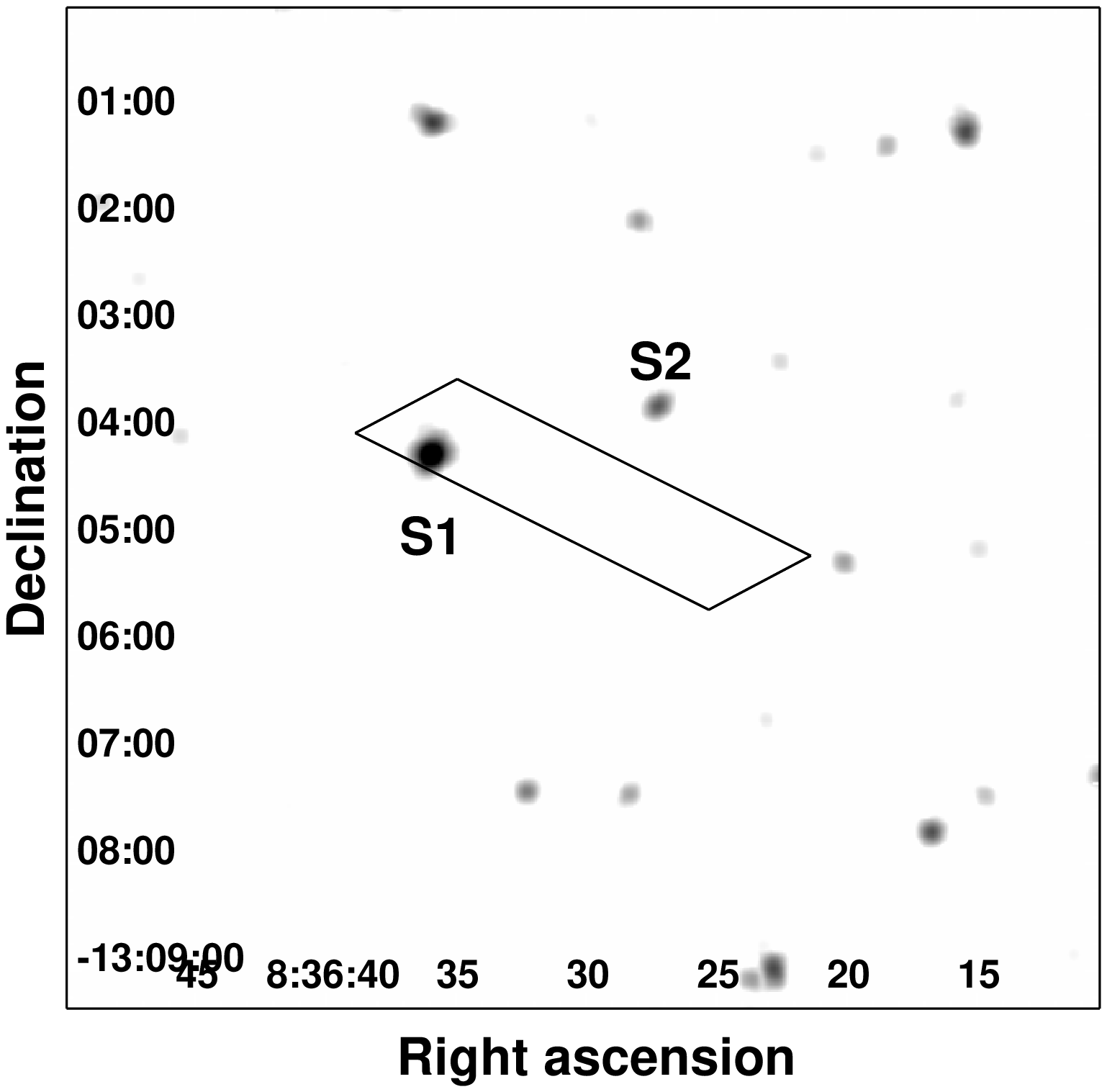}\\
\plotone{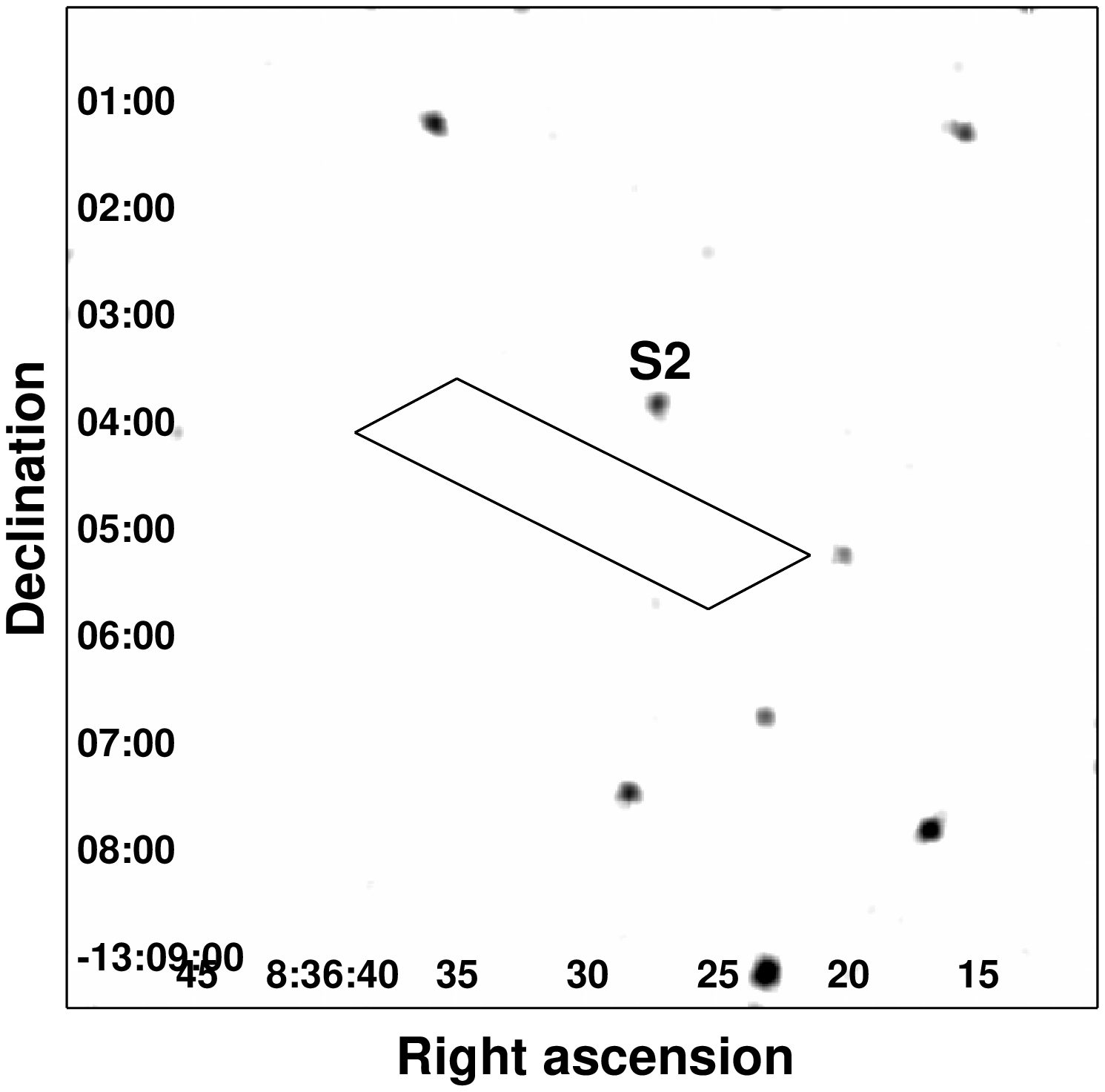}\\
\plotone{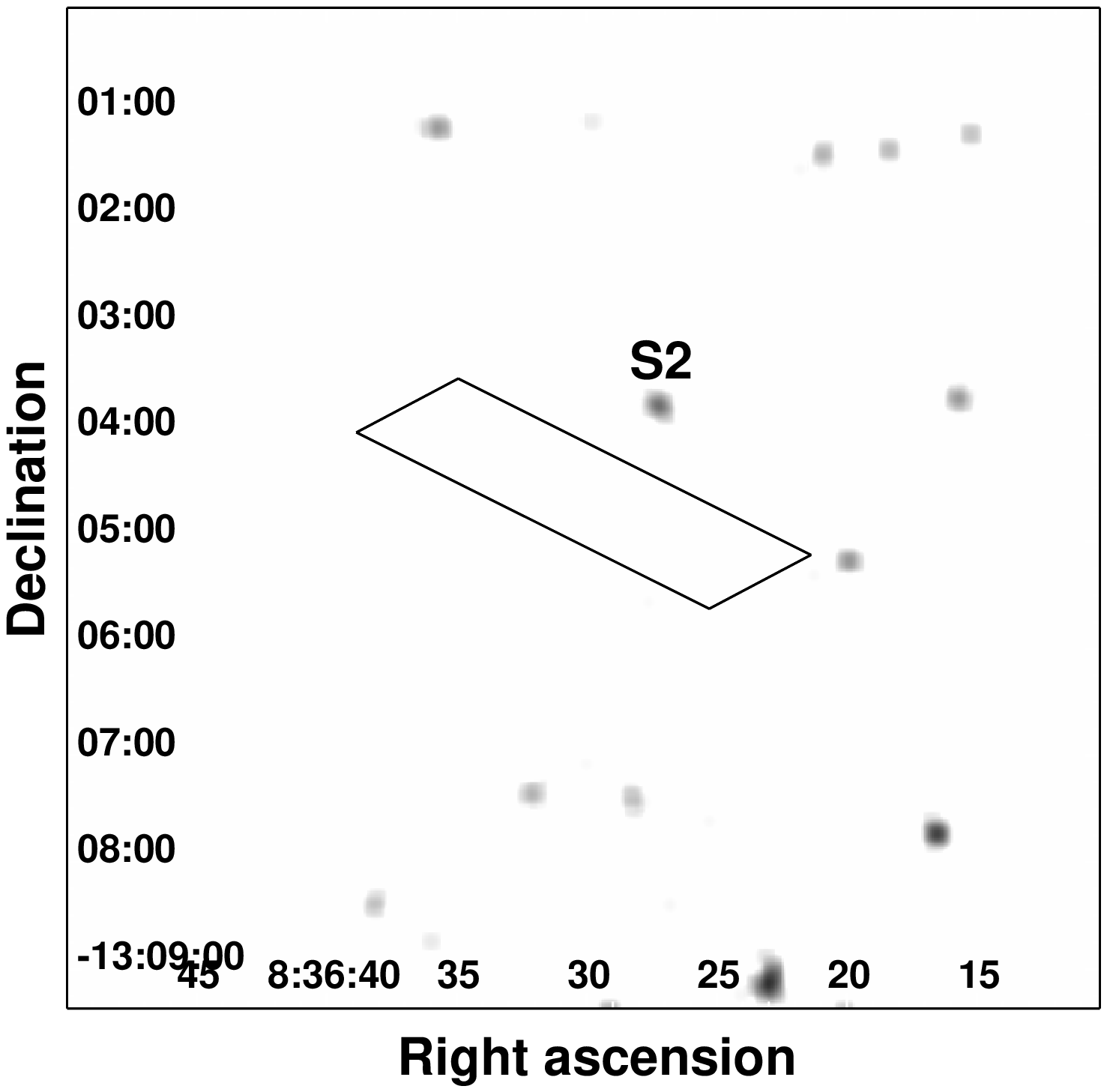}\\
\caption{{\it XMM-Newton} co-added pn+MOS1+MOS2 images showing the afterglow, 
S1, and the source S2, with the final RXTE-ASM/IPN error box overlaid, at three
epochs: Top: 1.88~days after the burst. Middle: 761~days after the burst.
Bottom: 910~days after the burst. 
See Table~\ref{xmmlog} for the exposure time of invididual images.}
\label{xmmim}
\end{figure}

We have used the final {\it Ulysses}, NEAR, and {\it Wind}-Konus data
to obtain a refined IPN error box for this burst.  The \it Ulysses - \rm Konus
annulus is centered at R.A.(2000)$=169{\fdg}8656$, 
decl.$(2000)=-56{\fdg}6466\rm$,
and has a radius of $53{\fdg}5052 \pm 0{\fdg}0073$ ($3 \sigma$).  The
{\it Ulysses} - NEAR annulus is centered at 
R.A.$(2000)=145{\fdg}4502$,
decl.(2000)$=-36{\fdg}9464$, and has a radius of 
$27{\fdg}9969 \pm 0{\fdg}0275$ 
$(3 \sigma)$.  The intersection of these annuli defines a 96~arcminute$^2$
error box.  However, the intersection of the {\it Ulysses}-Konus annulus
with the ASM error box defines a 3.3~arcminute$^2$ error box, whose
coordinates are given in Table~\ref{error}. The combined ASM-IPN error 
box is shown in Figure~\ref{xmmim}.

\begin{figure}
\epsscale{0.8}
\centering
\plotone{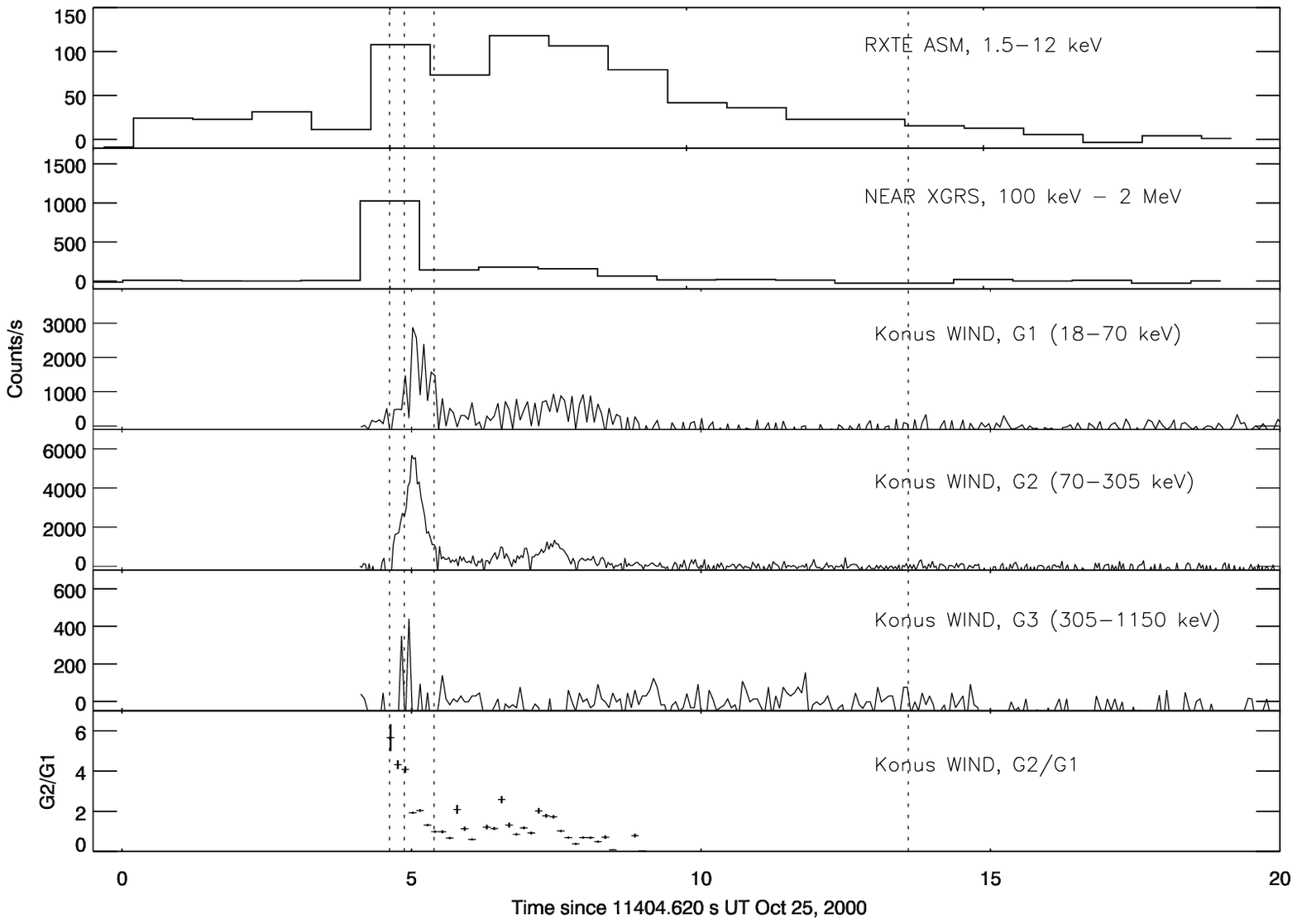}
\caption{Lightcurves of GRB~001025A. 
Plotted from top to bottom are the background subtracted lightcurves of 
RXTE ASM (1.5--12~keV, 1~s resolution), NEAR XGRS (100~keV--2~MeV, 
1~s resolution), {\it Wind}-Konus in three different energy bands 
(18--70~keV, 70--305~keV, 305--1150 keV, 0.128~s resolution), and 
the hardness ratio between the {\it Wind}-Konus 70--305~keV and 18--70~keV 
energy bands as function of time. The vertical dotted lines indicate
the intervals where time-resolved spectroscopy was done, see 
Fig~\ref{konus_spec}.
}
\label{all_time}
\end{figure}

\begin{figure}
\epsscale{0.8}
\plotone{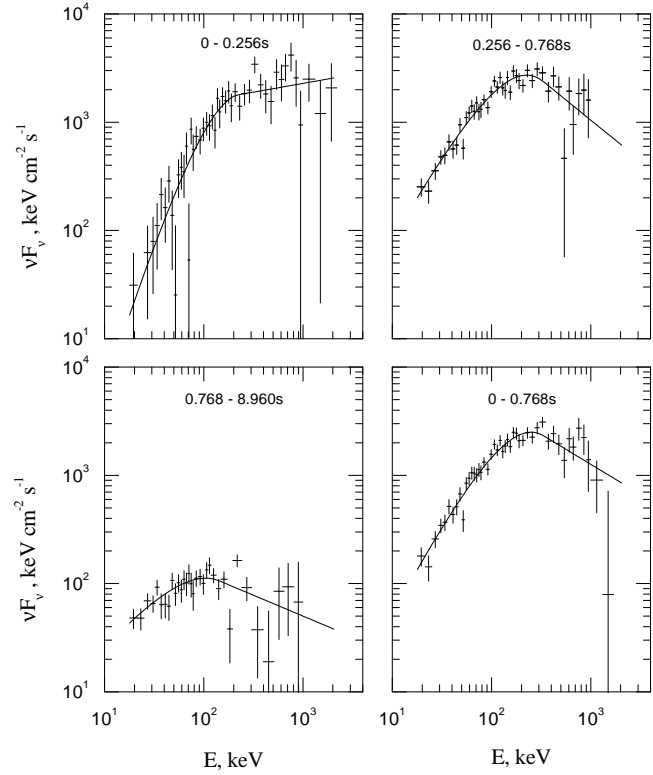}
\caption{Time-resolved energy spectra of GRB~001025A in $\rm \nu F_{\nu} $ 
units from {\it Wind}-Konus.  
Spectral fitting was done using the Band model. The fitting parameters are 
given in Table~\ref{konus_spec_tab}. 
The time intervals used for the spectral fits are indicated in 
Figure~\ref{all_time}.
}
\label{konus_spec}
\end{figure}

The ASM, Konus, and XGRS lightcurves are shown in 
Figure~\ref{all_time}. With a T$_{90}$ duration of 2.9 s (i.e., the
time to accumulate between 5\% and 95\% of the photons) in
the 50--200 keV energy range, this burst falls into the ``long-duration'' 
category.
Figure~\ref{all_time}
shows the lightcurve in several energy ranges, and indicates the intervals 
used for time-resolved spectral analysis. This figure also shows the hardness
ratio as a function of time.  A hard-to-soft evolution, which is 
often observed in GRBs \citep[e.g.][]{preece}, is evident.  
The energy spectra are presented in Figure~\ref{konus_spec}. 
The Band \citep{band} model, which consists of two smoothly joined
power laws, was used to perform spectral fits.  The Band function is:

\begin{equation}
f(E)=\left\{\begin{array}{ll}
A (E/100)^{\alpha} \exp{(-E/E_{\rm 0})}\nonumber & 
{\rm if} \quad E < (\alpha-\beta)E_{\rm 0}\\
A \{(\alpha-\beta) E_{\rm 0}/(100)\}^{(\alpha-\beta)}
\exp{(\beta-\alpha)} (E/100)^{\beta}\nonumber & 
{\rm if} \quad E \geq (\alpha-\beta)E_{\rm 0}\nonumber
\end{array}
\right.
\label{eqband}
\end{equation}

This yields a peak energy in the range $E_0\sim 85-140$~keV during the burst
(see Table~\ref{konus_spec_tab}). The 15--2000 keV peak flux of 
this burst was $ \rm 3.3 \times 10^{-5} \, erg \, cm^{-2} \, s^{-1} $ 
over 0.048~s, and the 15--2000 keV fluence was 
$ \rm 1.3 \times 10^{-5} \, erg \, cm^{-2} $.   
Neither the time history, nor the
spectrum, nor the intensity of this event was exceptional in any way.   
\citet{att} has proposed a simple redshift estimator based on the photon 
flux, energy spectrum, and duration of a GRB.  Using this method, 
we obtain $z\approx 0.8$, with a statistical uncertainty of about 50\%.

\subsection{Afterglow Searches}

In the weeks immediately following GRB~001025A, optical, X-ray, and radio 
observations were carried out in attempts to identify an afterglow. 

An optical afterglow was not identified, either in observations starting 
1.16~days after the burst \citep[][$R>22.5$]{fynbo}, or in observations starting 
1.66~days after the burst \citep[][$R>18.0$]{uemura}.

An {\it XMM-Newton} observation $\sim 2$~days after
the burst revealed one apparently fading source (S1, see Figure~\ref{xmmim}) 
in the initial ASM/IPN 
error box \citep{alta}. In late-time follow-up {\it XMM-Newton} observations 
this source had disappeared, strongly indicating that (S1) is the X-ray 
afterglow of GRB~001025A.

Radio observations with the {\it Very Large Array} (VLA) were carried out
at two epochs, 
but they did not reveal any radio afterglow. The first VLA observation
took place November~1, 2000, approximately 6 days after the GRB 
(D.~Frail 2003, private 
communication). Unfortunately, the {\it XMM-Newton} attitude was
subsequently revised \citep{altb}, and the refined position of
S1 was $\rm \sim 2 \arcmin$ from the initial one. 
Due to the off-set between the VLA pointing and the revised position of
S1, the VLA sensitivity at the position of S1 was relatively low. 
VLA observations took place again on November~21 and 24, 2000 
(27--30 days after the burst), but given the timescales and sensitivies
($\rm \sim 200 \mu Jy$) most GRB radio afterglows would not have been detected.
Thus it is not clear whether this burst was radio-quiet.  

In the following three sections, we present our analysis and results of 
optical observations and X-ray observations of the GRB~001025A afterglow.

\subsection{Optical Observatons with the VLT}

 The field of GRB~001025A was observed at the {\it Very Large Telescope}
(VLT) equipped with the FORS1 camera at three epochs, 
under photometric and good seeing conditions (see Table~\ref{vltlog} for 
details). 
The first epoch was a few hours after the release of the IPN error box 
(Hurley et al. 2000), the second epoch was 24~hours later, and the 
final epoch was 5 months later.

\begin{figure}
\epsscale{0.8}
\plotone{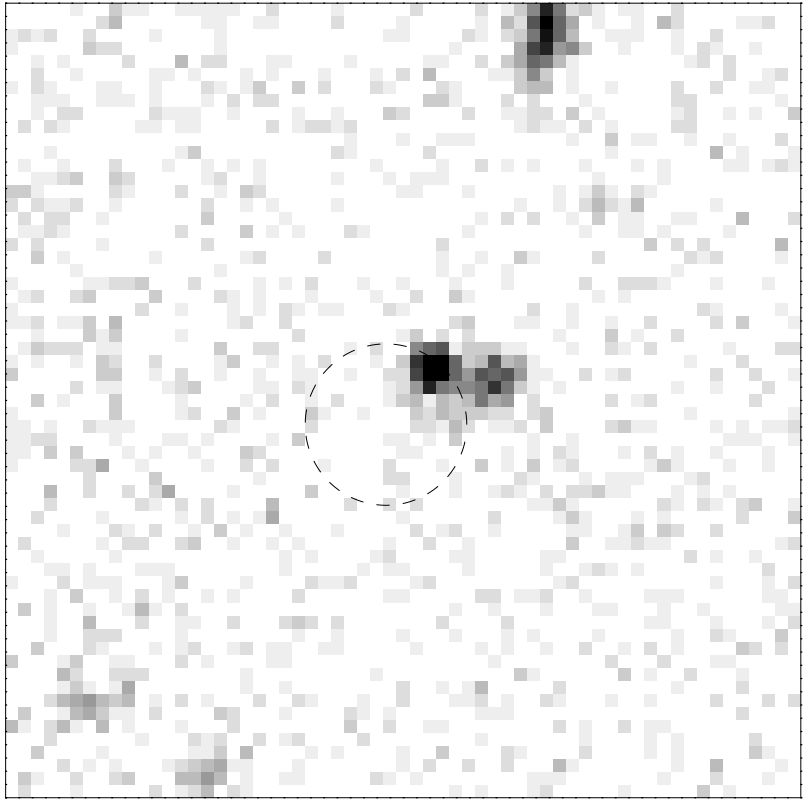}
\caption{$12 \times 12$~arcsec$^2$ VLT $R$-band image with the $2\sigma$ 
error circle of source S1. North is up and East is to the left.
}
\label{host}
\end{figure}

By comparing the combined $R$-band images of the three epochs we find no 
transient sources in the IPN error box and in particular near the 
{\it XMM-Newton} error circle of S1 to a 3$\sigma$
detection limit of $R=25.5$. Figure~\ref{host} shows a 12$\times$12
arcsec$^2$ section of the last epoch VLT image with the {\it XMM-Newton} 
2$\sigma$ 
error circle overplotted. The 2$\sigma$ error circle is consistent
with the eastern part of what seems to be either a galaxy with two 
main components or a random projection of two unrelated objects.
Following \citet{piro}, we estimate that the probability for a
chance alignment is less than 1\%, so we consider this object the likely
host galaxy of GRB~001025A.
The total magnitude of this complex is $R=24.01\pm$0.04 and 
$B=25.13\pm$0.12 \citep[corrected for a modest foreground extinction of 
$E(B-V)=0.07$][]{schlegel}.

\begin{figure}
\epsscale{0.8}
\plotone{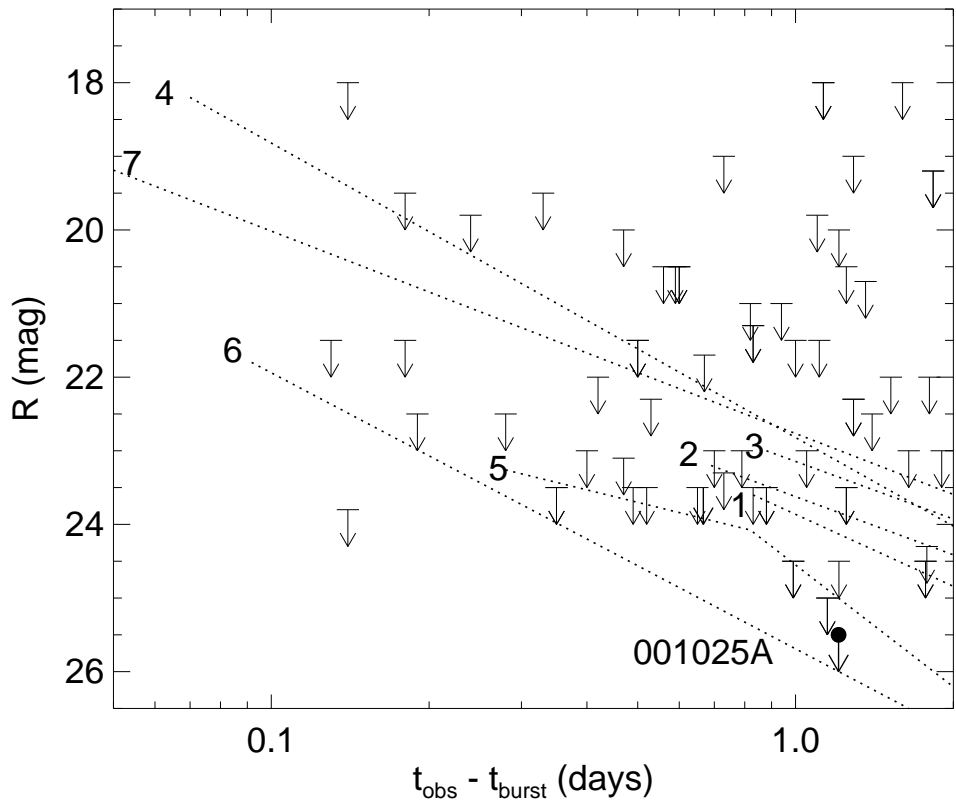}
\caption{$R$-band magnitudes and upper limits as a function of time 
for various gamma-ray bursts in the pre-{\it Swift} era, 
compared to the observations of GRB~001025A (filled circle).  
The dotted lines sketch the decay of optically dim bursts, starting at the
time of the first observation. The labels are as follows:
(1) GRB~980329, (2) GRB~980613,
(3) GRB~000630, (4) GRB~020124, (5) GRB~020322, (6) GRB~030115, 
(7) GRB~021211.
}
\label{dark}
\end{figure}

GRB afterglows display a wide variety of fading behaviors 
(see Figure~\ref{dark}).
Of the pre-{\it Swift} detected optical afterglows, all but GRB~030115
would have been detected in the VLT observations. The failure to detect the
optical afterglow of GRB~001025A is thus not because our observations are
less sensitive than studies of other detected GRB afterglows.

\subsection{X-ray Observations with {\it XMM-Newton}}

{\it XMM-Newton} carried out observations of the region around the error 
box at three different epochs, spanning 1.9--2.3 days, 761.0--761.2 days, and  
910.2--910.4 days after the GRB (see Table~\ref{xmmlog}). 
The {\it XMM-Newton} target of opportunity observation $\sim 2$~days after
the burst revealed at least four X-ray sources in the initial ASM/IPN error box 
\citep{alta}, one of which (S1) appeared to be fading.  
As shown in Figure~\ref{xmmim}, only S1 is in the 
final ASM/IPN error box. This, along with marginal evidence for a fading behavior 
\citep[see Section 4 below, and][]{watson},
suggested that it was the X-ray afterglow of GRB~001025A.
In order to determine with certainty whether S1 was indeed the fading afterglow 
(as
opposed to a random, variable X-ray source such as an active galactic nucleus),
we carried out a follow-up {\it XMM-Newton} observation.  This
observation started on November~25, 2002, but had to be terminated because of
high background radiation; it was completed on April 23, 2003.

The first observation has already been reported on in \citet{watson}, 
but for the analysis presented here, we have made two improvements:
we have omitted data taken during periods with
high background, and we have analyzed data from the second and third
epochs as well. 
The effective EPIC pn exposure times of the three {\it XMM-Newton} observations, 
after
screening out high background intervals, are 15~ks, 9~ks, and 18~ks
respectively.

The data were processed and analysed with SAS version 5.4.1 and XSPEC version
11.3. Periods with flares were identified based on the
10--12~keV lightcurve of the full EPIC detector and events registered
during these periods (where the pn count rate 
was above 1.5 counts/s, and where the MOS count rate was
above 0.25 counts/s) were filtered out. For the pn detector, single and
double events were included in the analysis and the two exposures taken
in the first epoch were co-added. For the MOS detectors, single, double,
triple, and quadruple events were included in the analysis and the
two MOS data sets for each epoch were co-added.
pn events in the energy interval 0.3--9~keV 
and MOS events in the energy interval 0.4--7~keV  were included in the analysis.
Spectra and lightcurves of sources were
extracted using a circular aperture of radius 18\arcsecs.
Background events were extracted from an
annulus centered on the source in question with an inner radius of 
18\arcsecs{} and an outer radius of 50\arcsecs. 
The background regions are entirely on the same
chip as the relevant sources, and they do not include other contaminating sources.

Below we present results from our analysis of the {\it XMM-Newton} images, 
lightcurves, and spectra of the GRB~001025A X-ray afterglow.

\subsubsection{Imaging}
In the first epoch observation two sources were detected, S1 
R.A.(2000)$= 8{\hours}36{\mins}35{\fs}93$, 
decl.(2000)$=-13{\degr}04{\arcmin}10{\farcs}82$, 
$\pm 0{\farcs}3$), 
and S2, in and in the vicinity of the final ASM/IPN error box, respectively
(see Figure~\ref{xmmim}).
Over the first {\it XMM-Newton} observation 264 net counts (0.3--12~keV) 
were detected for source S1 by the pn detector, corresponding to a flux of 
$(6.4 \pm 0.2)\times 10^{-14}$~erg~cm$^{-2}$~s$^{-1}$.
For source S2, 78 net counts (0.3--12~keV) were detected, corresponding to a 
flux of $(2.0\pm 0.4)\times 10^{-14}$~erg~cm$^{-2}$~s$^{-1}$.
By the time of the second epoch observation, S1 had faded beyond detection 
($f_X$(0.3--12~keV)$<6\times 10^{-15}$~erg~cm$^{-2}$~s$^{-1}$)
while the flux of S2 was consistent with that of the first epoch. 
In the third epoch observation S1 was still undetectable 
($f_X$(0.3--12~keV)$<5\times 10^{-15}$~erg~cm$^{-2}$~s$^{-1}$)
and the flux of S2 was consistent with the flux 
at the first two epochs. Hence, source S2 is persistent over several years,
displaying flux variations less than about 25\% from epoch to epoch.
S2 is coincident with an $R\sim 20$ optical source which based on a spectrum
obtained with the {\it Nordic Optical Telescope} seems to be
a QSO at redshift $z\approx2.5$ (M.~I.~Andersen, private 
communication).

\subsubsection{Lightcurve}

The pn lightcurve for S1 is well fitted by a power law with decay index 
$\delta_X =-1.6 \pm 0.35$ ($\chi^2 =4.80$ for 8 d.o.f., 0.3-12~keV). 
This is somewhat lower than, but consistent with, the steep decay index 
$\delta_X =-3.0 \pm 1.9$ derived by \citet{watson}. The data quality of the
corresponding co-added MOS1+2 lightcurve does not allow us to further constrain
the fading of S1. The decay of S1 is typical of X-ray afterglows having a decay 
index of $\delta_X \approx -1.4$ \citep{piro01}.
By extrapolating the best-fit pn lightcurve we predict the total number 
of counts in the second and third epoch {\it XMM-Newton} observations 
from S1 to be 0.04 counts and 0.02 counts, respectively (i.e., well below 
detectability).
Hence, the second and third epoch observations establish
that S1 is a non-persistent source; hence we identify S1 as the X-ray 
afterglow of GRB~001025A.

\begin{figure}
\epsscale{0.8}
\plotone{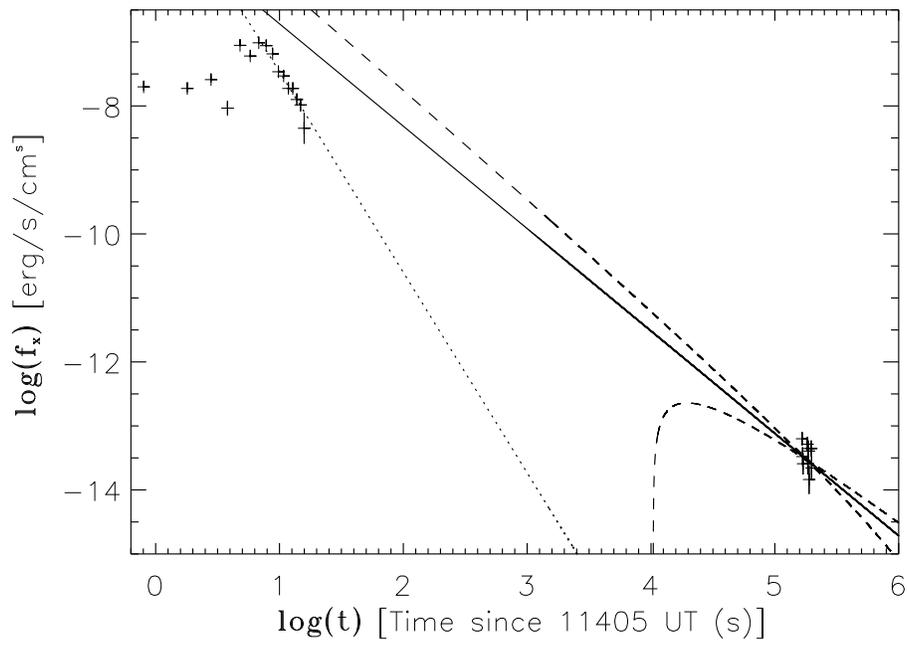}
\caption{RXTE ASM and {\it XMM-Newton} pn 1.5--12~keV lightcurves
with power law fits overlaid. Dotted line: Fit to the tail of the RXTE ASM 
lightcurve. Full line: Fit to {\it XMM-Newton} pn 0.3--12~keV lightcurve 
renormalized to  the 1.5--12~keV flux level.
Dashed lines: $1\sigma$ envelope of fit to {\it XMM-Newton} pn 0.3--12~keV
lightcurve renormalized to  the 1.5--12~keV flux level.
}
\label{xmm_lc}
\end{figure}

In order to compare the X-ray afterglow to the prompt X-ray emission
as observed by the RXTE ASM we extracted the pn 1.5--12~keV lightcurve. 
A power law fit to the 1.5--12~keV light 
curve is not well constrained (decay index $\sim -1$). Hence, we extrapolated
the best fit power law to the 0.3--12~keV light curve (renormalized to the
1.5--12~keV flux level) to the prompt emission phase. This extrapolation
reproduces the RXTE ASM flux within the uncertainties (see Figure~\ref{xmm_lc}).
However, fitting a power law to the fading of the prompt emission in the 
1.5--12~keV band yields a steep power law (decay index $-3.1\pm 0.4$) 
underestimating the pn data points. Such a steep decay has been reported
in several early X-ray afterglows observed with the {\it Swift} X-ray Telescope 
\citep{tag}. It is not possible to fit a power law decay jointly to the fading prompt 
emission and the X-ray afterglow.
The X-ray afterglow observed by {\it XMM-Newton} is therefore not a simple 
continuation of the prompt emission.

\subsubsection{Spectral Analysis}
Spectra of S1 were extracted from the first epoch observations and binned 
with a minimum of 20 counts per bin. Several spectral models were fitted to
the pn and MOS1+2 spectra simultaneously: (i) power law models, representing 
synchrotron emission from a population of relativistic electrons in the 
fireball \citep{spn}, 
(ii) thermal plasma emission giving rise to X-ray emission lines as seen
in several other GRB afterglows \citep{reeves,watson03} 
\citep[the MEKAL model as is typically used to represent thermal emission][]{mewe,lied}, 
and (iii) combinations of (i) and (ii). In all models Galactic absorption was included,
and additional absorption at the burst redshift was included in some of the models. 
A summary of the spectral fits is given in 
Tables~\ref{PL},\ref{mekal},\ref{PLmekal} with errors quoted as 90\% 
confidence intervals. 

An absorbed synchrotron model is an acceptable fit to the spectrum 
(see Table~\ref{PL}), but the best-fit 
photon index is rather steep 
\citep[$\Gamma \sim 2.8$, consistent with the findings of][]{watson}.
The Galactic \ion{H}{1} column density in the direction of GRB~001025A is 
$6.1\times 10^{20}$~cm$^{-2}$ and can in worst case be uncertain by
up to 50\% \citep{dicklock}. Power law models indicate absorption in excess
of Galactic absorption, but the column density of extra-galactic absorbing
material is not strongly constrained. In this model the 
burst redshift is unconstrained.

An absorbed thermal plasma model (MEKAL) is as good a fit as the absorbed
synchrotron emission model (see Table~\ref{mekal})
and implies a plasma temperature around 3~keV and 
a burst redshift around $z\sim 0.3$. \cite{watson} favor a redshift in the range 
0.5--1.2 from their MEKAL model fits, and we note that the thermal plasma could
be outflowing resulting in a host galaxy redshift lower by $\Delta z \sim 0.1$ 
\citep[e.g.][]{reeves}. Hence, thermal emission models do not provide strong 
constraints on the burst redshift.
The absorbing column density at the burst redshift 
is at most twice the Galactic column density. 
The plasma abundances are 
not well constrained, but they are consistent with the solar values 
($Z_{Fe}\lesssim 2 Z_{\odot}$, $Z_{S,Si}\lesssim 13 Z_{\odot}$).

\begin{figure}
\includegraphics[angle=-90,scale=.7]{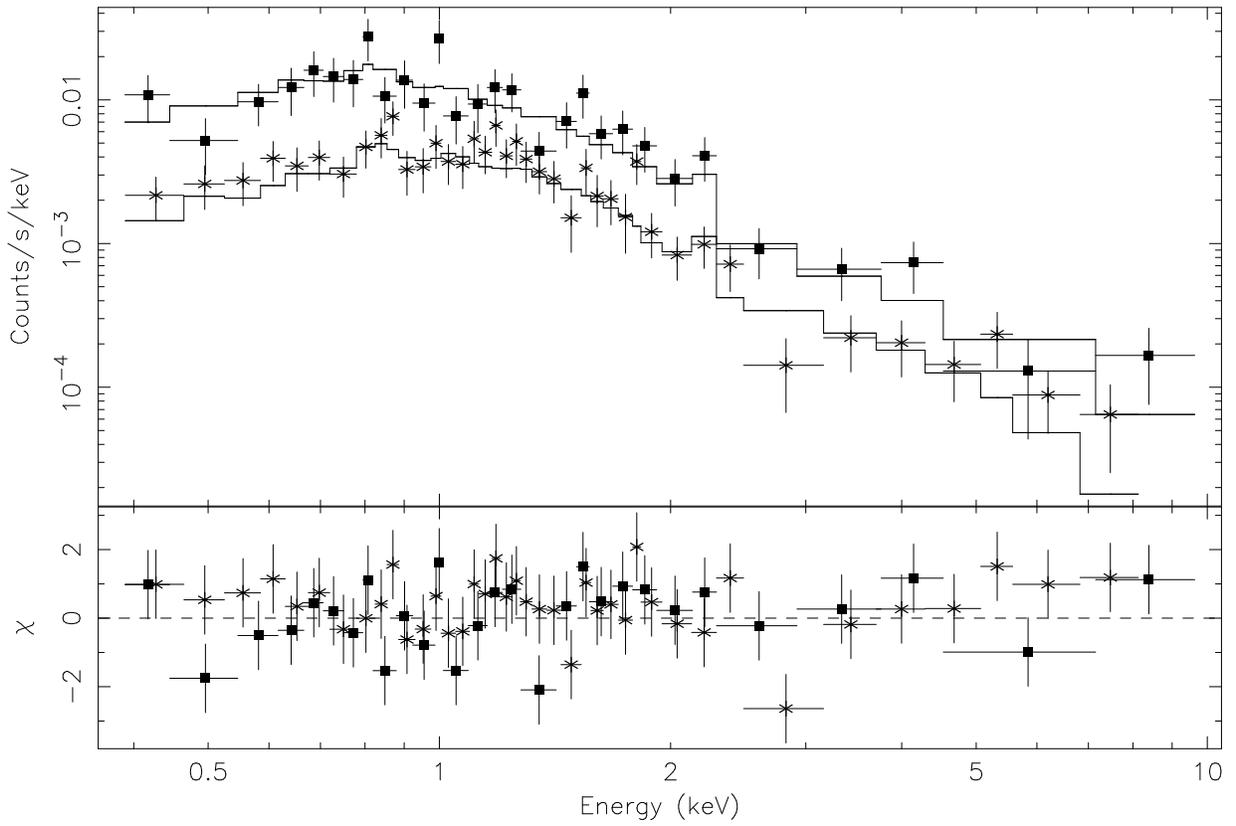}
\caption{Top panel: {\it XMM-Newton} pn (squares) and MOS1+2 (stars)
spectra of the afterglow of GRB~001025A with a spectral binning of min. 10
counts per bin. Overplotted is the best-fit combined synchrotron/thermal
emission model with fixed Galactic absorption and absorption at the
burst redshift. Lower panel: residuals of model fit.
}

\label{xmmspec}
\end{figure}

Naturally, a model with a combination of a synchrotron emission component and a thermal 
emission component (see Figure~\ref{xmmspec}) is also a good fit 
(see Table~\ref{PLmekal}).
The relative normalization of the synchrotron emission component and the
thermal emission component is not well constrained, but the
synchrotron emission component contributes $\sim 2/3$ of the total flux
and above $\sim 2$~keV the synchrotron emission component dominates.
The best-fit photon index is lower than in the pure synchrotron emission model 
fits ($\Gamma \sim 1.8$), 
but close to typical photon indices for X-ray afterglows \citep{piro01}. 
The favored burst redshift is $z \approx 2$, and there is an indication of
circumburst absorbtion ($N_H\sim 10-20\times 10^{21}$~cm$^{-2}$).
The abundances of the thermal plasma are not well constrained,
but they are consistent with the solar values
($Z_{Fe,S,Si}\lesssim 6 Z_{\odot}$).

In conclusion, an absorbed synchrotron model, an absorbed thermal plasma model,
and a combination of the two all represent a good fit to the data. Consequently,
we consider all these models in the following.

\subsection{Spectral Energy Distribution}
\begin{figure}
\epsscale{0.8}
\plotone{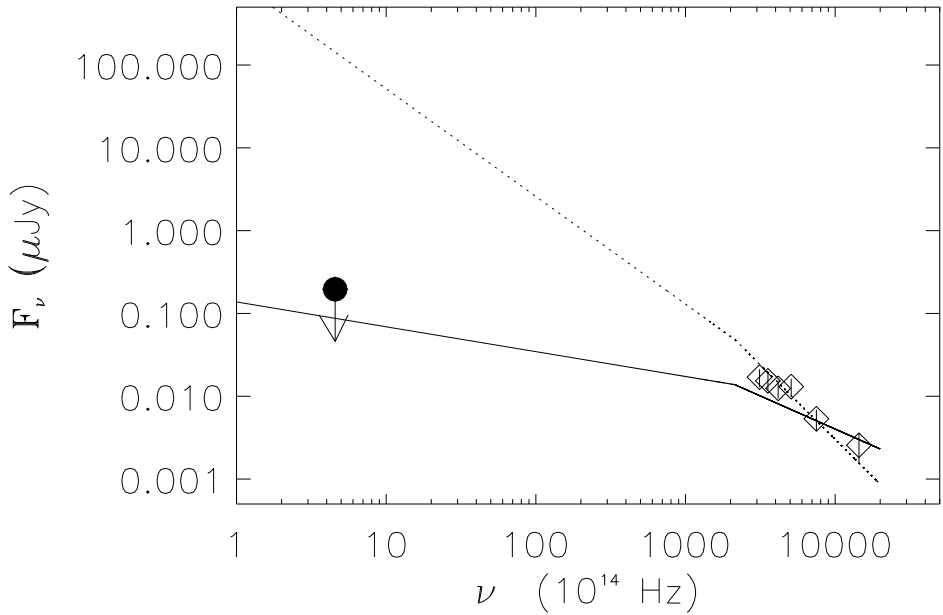}
\caption{The optical to X-ray spectral energy distribution of GRB~001025A 1.20~days
after the burst.
The {\it XMM-Newton} pn spectrum recast to this epoch is shown as diamonds (only
data points above 1.3~keV, where absorption is neglibible), the $R$-band upper limit 
is shown as the 
filled circle. Overlaid is the synchrotron emission component of the best-fit from 
the combined synchrotron emission/thermal emission model (spectral index 0.8, full line) 
and the best-fit synchrotron emission model (spectral index 1.8, dotted line).
The extrapolation to the optical band assumes that the cooling break is at
1~keV ($2140 \times 10^{14}$~Hz) and that the spectral index changes by 
$-0.5$ \citep{spn} when going to the short frequency side of the break. 
}

\label{sed}
\end{figure}

We derived the broadband spectral energy distribution 
by extrapolating the best-fit X-ray spectral models 
to the optical band. We used the X-ray decay index $\delta_X =-1.6$ to recast 
the X-ray spectrum to the epoch of the first $R$-band observations (1.20 days
after the burst). For the synchrotron emission model (spectral index $\beta_X=1.8$) 
an observed $R$-band extinction of at least 7~magnitudes is
required to bring the model in agreement with the $R$-band upper limit
(see Figure~\ref{sed}).
On the other hand, extrapolating the synchrotron emission component from the best-fit 
combined synchrotron emission/thermal emission model (spectral index $\beta_X=0.8$) 
to the $R$-band shows that the $R$-band upper limit is consistent 
with the model, if the cooling break is close to the X-ray band.

\section{DISCUSSION}

The spectral index of the (X-ray) afterglow is a key parameter for constraining
the properties of the burst and its environment. 
The pure synchrotron emission fit to the GRB~001025A X-ray afterglow has a
very steep spectral index for a GRB X-ray afterglow.
On the other hand, the synchrotron component of the combined thermal 
plasma/synchrotron model has a spectral 
index in line with what is found for other GRBs \citep{piro01}.
Detection of thermal afterglow emission has been reported in several other GRBs
\citep[see][for an account of GRBs with thermal emission]{watson03}
increasing the soft X-ray flux in the $\sim 0.5-2$~keV range.
One of the most prominent claims of thermal emission is in
the GRB~011211 X-ray afterglow where thermal emission is
detected during the first 5~ks of the {\it XMM-Newton} observation \citep{reeves}. 
We investigated the effect on the spectral index of fitting an
absorbed synchrotron model to a spectrum with thermal emission 
by fitting an absorbed synchrotron emission model to the first 5~ks of the GRB~011211
{\it XMM-Newton} spectrum.
We obtain a spectral index of $1.6\pm 0.3$; this is very similar to the spectral 
index ($\beta_X=1.8\pm 0.3$) we find fitting the same model to GRB~001025A.
The steep spectral index of a pure synchrotron emission fit to GRB~001025A
compared to the shallower index from a synchrotron emission plus thermal
emission fit thus suggests that thermal emission may be 
steepening the X-ray afterglow spectrum of GRB~001025A. 

Due to the different best fit spectral indices of the synchrotron emission in the 
pure synchroton model and in the combined thermal plasma/synchrotron model,
these two models have quite different implications for the deduced properties of
the GRB~001025A fireball and its environment.

\subsection{Physics of the Fireball}
The prompt soft X-ray emission has a steep decay as seen in the 
tails of other GRBs \citep{smith02,giblin}. 
Observations with the {\it Swift}~XRT has revealed a population of
bursts with an early steep decay and a shallower late time decay \citep{tag}
similar to the behavior of GRB~001025A.

Several models predicts an early steep X-ray decay, including a hot cocoon 
surrounding a relativistic jet \citep{rr}, emission from outside the 
relativistic beaming cone, $\theta > \Gamma^{-1}$, \citep{kp}, and
a single spherical shell emitting instantaneously \citep{fen}.
The steepness of the prompt emission decay 
excludes that the tail of the prompt emission is the onset of 
the afterglow \citep{spn}. However, the afterglow could appear at any time
$>16$~s after the burst at which epoch the GRB~001025A afterglow 
is expected to be below the sensitivity of RXTE ASM.

The X-ray afterglow has a steep spectral index, if it is due to synchrotron
emission only. The standard fireball model predicts in this case a temporal
decay index of $-2.2\pm 0.45$~to~$-3.2\pm 0.45$ for a spherical blast wave
and an index of $-3.6\pm 0.6$~to~$-4.6\pm 0.6$ for a jet \citep{zhangmes}.
This is steeper than observed ($\delta_X=-1.6\pm 0.35$),
but only the jet model can be ruled out.
If the X-ray afterglow is due to a combination of thermal plasma and synchrotron 
emission, the fireball model predicts a shallower afterglow decay index 
of the synchrotron component \citep[index $-0.4\pm 0.3$~to~$ -1.7\pm 0.9$ for a 
spherical blast wave, index $-1.9\pm 0.3$~to~$-2.6\pm 1.2$ for a jet,][]{zhangmes} 
more in line with the X-ray lightcurve. In this case
models where the X-ray afterglow originates from fast cooling are ruled out.

\subsection{Burst Redshift and Luminosity}
In the combined synchroton/thermal plasma spectral model of the X-ray 
afterglow, the suggested burst redshift is $z \approx 2$. 
This yields an isotropic prompt energy release of $1.5\times 10^{53}$~erg
(for a $H_0=72$~km/s/Mpc, $\Omega_m=0.3$, $\Omega_{\Lambda}=0.7$ cosmology)
which is quite typical \citep[e.g.][]{bloom}.
The isotropic 0.3--12~keV afterglow luminosity is 
$L_X=(3.2 \pm 2.6)\times 10^{45}$~erg/s 
time-averaged over the first {\it XMM-Newton} observation.
We now compare the X-ray luminosity of GRB~001025A
to the luminosities of the GRB sample of \citet{berger03}.
We use our derived X-ray afterglow decay index, flux and spectral 
index to calculate the isotropic luminosity 
\citep[$L_{X,iso}$, eq.~(1) in][using their cosmology]{berger03}
of GRB~001025A 10~hours after the burst. 
In the combined synchrotron/thermal plasma
model with a burst redshift $z=2$, we find 
$L_{X,iso}=(2.4 \pm 1.9)\times 10^{46}$~erg/s.
This is close to the typical X-ray afterglow luminosity at this epoch for the 
\citet{berger03} burst sample. 
However, in that sample, redshifts are not known for most of the bursts,
in which case the luminosity has been derived assuming a redshift of $z=1.1$.
If GRB~001025A is at a redshift of $z=1.1$ its isotropic luminosity is 
$L_{X,iso}=(3.5 \pm 2.5)\times 10^{45}$~erg/s which is in
the low tail of the \citet{berger03} X-ray luminosity distribution.
If the burst redshift is $z=0.33$, as predicted in the thermal plasma only 
model,
the luminosity of GRB~001025A $L_{X,iso}=(2.3 \pm 1.5)\times 10^{44}$~erg/s
is lower than all GRBs in the \citet{berger03} sample.

The $B$-band detection of the likely host galaxy implies a redshift $z<2.7$.
The magnitude and color of the galaxy are consistent with a
redshift around two, but they do not constitute strong constraints on the redshift.

The pseudo-redshift estimate \citep{att} for this burst $z=0.8\pm 0.4$,
and although not strongly constraining, this is in line with the lower redshift 
derived from the thermal emission models of the X-ray afterglow.

\subsection{The Burst Environment}
The appearance of the candidate host galaxy is similar to other GRB
hosts, i.e., an irregular morphology and a color that for a redshift of
$z>0.3$ is indicative of active star formation.
From the absorbing column density in the soft X-ray band, 
the extinction towards the burst can be estimated by assuming a 
gas-to-dust ratio. The best-fit column densities
at the burst redshift are $N_H\lesssim 20\times 10^{21}$~cm$^{-2}$. 
Taking a Galactic gas-to-dust ratio \citep{ps}, 
this translates into a restframe extinction of $A_V\lesssim 11$ for $z<2$. 
Using an SMC gas-to-dust ratio results in $A_V\lesssim 1$, i.e.\ not enough
to account for the extinction required in the pure synchrotron model for
the X-ray afterglow.
A Galactic gas-to-dust ratio is at variance with the findings for GRBs with 
detected optical afterglows generally having SMC-like gas-to-dust ratios \citep{galwij}.
Either the pure synchrotron model is not appropriate for the GRB~001025A
X-ray afterglow or GRB~001025A is much more obscured than bursts
with detected optical afterglows 
\citep[typically having $A_V\lesssim 0.2$][]{hjorth03a}.

\subsection{What Makes GRB~001025A a Dark Burst?}
The dark bursts in the GRB sample of \citet{pall} 
\citep[updated as of July 8, 2005,][]{pallhtml} has a mean 1.6--10~keV flux of
$3.0 \times 10^{-12}$~erg~cm$^{-2}$~s$^{-1}$ 11~hours after the 
burst. Extrapolating the best-fit X-ray lightcurve of GRB~001025A
to this epoch, the estimated 1.6--10~keV flux is 
$5.0\pm 2.7\times 10^{-13}$~erg~cm$^{-2}$~s$^{-1}$. The X-ray
afterglow of GRB~001025A is fainter than any of the dark bursts in the \citet{pall}
sample and much fainter than the mean of GRBs with detected optical afterglows
in this sample ($4.2 \times 10^{-12}$~erg~cm$^{-2}$~s$^{-1}$).

The low $R$-band flux of GRB~001025A is not caused solely by a general faintness
of the afterglow at all wavelengths.   
The $R$-band-to-3~keV spectral index is $\lesssim 0.43$
so GRB~001025A is clearly dark according to the definition of \cite{pall}.
For GRBs with detected optical afterglows, 
the $R$-band-to-3~keV spectral index is $0.73$ and for {\it Swift} detected bursts 
it is $0.65$ \citep{pallhtml} so the afterglow of GRB~001025A is
optically faint relative to the X-ray band. 
The low optical-to-X-ray spectral index could be due to heavy obscuration 
in the burst environment. Alternatively, the cooling frequency could be close to 
the X-ray band, giving rise to a shallow spectral index requiring no optical 
obscuration.

The X-ray spectral index has been measured only for a few dark GRBs.
By re-normalizing the X-ray spectrum and the optical observations to a common
epoch (e.g. 11~h after the burst) and extrapolating the X-ray spectral index to 
the optical band, assuming a position of the cooling break, the $R$-band flux 
can be predicted.
For three of the five GRBs with measured spectral index in the dark GRB sample 
of \citet{depas} (GRB~990704, GRB~990806, GRB~000214) the predicted
$R$-band flux is lower than the observational upper limit, when invoking a 
cooling break at 1~keV. These three GRBs have steep X-ray spectral indices, 
possibly reflecting the presence of a thermal component on top of the synchrotron
emission. The true synchrotron spectral index may thus be lower than
the index obtained from a pure synchrotron emission fit, giving rise to even lower 
predictions for the optical flux. For the remaining two GRBs in the dark GRB sample 
of \citet{depas} (GRB~000210 and GRB~001109) obscuration is 
required in order to bring the observational upper limit in line with the model 
prediction (or the derived X-ray spectral index is too step due to the presence of
thermal emission). 
Two of the bursts detected by {\it Swift} and with published X-ray spectral
index \citep[GRB~050401 and GRB~050408,][]{chin} are dark according to the 
definition of \cite{pall}. 
For both bursts obscuration is required in order to be consistent with 
the upper limit on the $R$-band flux, even when invoking a cooling break at 1~keV.

\section{CONCLUSIONS}

GRB~001025A was a long-duration GRB with a prompt emission spectrum 
well fitted by the Band model. The tail of the prompt soft X-ray emission
decays too steeply to be the beginning of the afterglow.
Three epoch {\it XMM-Newton} observations show that the 
GRB~001025A afterglow is X-ray faint. 
Furthermore, the optical flux is relatively low compared to the X-ray band.
Either (i) the optical afterglow suffers extinction of at least seven magnitudes,
or (ii) a significant fraction of the X-ray afterglow flux originates from
a thermal plasma, and the cooling frequency is close to the X-ray band
at 1.2~days after the burst.
In the latter case, we predict that the burst redshift is around two,
and we find that the X-ray luminosity and the spectral index are fairly typical
of GRB afterglows.
Alternatively, the GRB~001025A X-ray afterglow spectrum is characterized by
a very steep synchrotron index, and GRB~001025A is situated in
an environment with an unusually large gas-to-dust ratio, consistent with
the Galactic ratio.

The present study of GRB~001025A, and the properties of the few other dark
GRBs with measured spectral index, raises the possibility that some
bursts appear optically dark because their afterglow
is faint, and their cooling frequency is close to the X-ray band.
The viability of this scenario can be tested from early observations 
of X-ray afterglows from dark GRBs yielding the temporal and spectral 
behaviour of the X-ray afterglows, which in turn constrain the geometry of the 
blast wave and the underlying physical processes.
This would provide a crucial test for physical differences
between dark bursts and bursts with an optical afterglow.

\acknowledgments

We are grateful to Dale Frail for making the results of his VLA observations
known to us. We thank the anonymous referee for useful and very detailed 
comments that improved 
the presentation of our results. KH is grateful for \it Ulysses \rm support 
under JPL Contract 958056, for support from the Long Term Space Astrophysics program 
under NAG5-3500, 
and for support as a Participating Scientist in the NEAR mission under NAG5-9503.
KP and JPUF acknowledge support from the Carlsberg foundation. KP acknowledges
support from Instrument Center for Danish Astrophysics.
PJ acknowledges support from a special grant from the Icelandic
Research Council. This work was supported by the Danish  
Natural Science Research Council (SNF). 
From the Russian side this work was suported by
the Russian Space Agency contract and RBRF grant 03-02-17517.
DAS is supported by an NSF Astronomy and Astrophysics Postdoctoral
Fellowship under award AST-0105221.
The authors acknowledge 
benefits from collaboration within the EU FP5 Research Training 
Network "Gamma-Ray Bursts: An Enigma and a Tool".
Based on observations obtained with {\it XMM-Newton},
an ESA science mission with  instruments and
contributions directly funded by ESA Member
States and NASA,and based on observations made with ESO Telescopes at the 
La Silla or Paranal Observatories under programme 66.A.-0386(A).

Facilities: \facility{XMM(EPIC)}, \facility{VLT(ANTU)}, \facility{RXTE(ASM)},
\facility{NEAR(XGRS)}, \facility{Wind(Konus)}.

\clearpage

\clearpage

\clearpage

\clearpage

\clearpage

\clearpage

\clearpage

\clearpage 

\begin{deluxetable}{cc}
\tabletypesize{\footnotesize}
\tablecaption{\it Corners of the final ASM-IPN Error Box for GRB~001025A}
\tablewidth{0pt}
\tablehead{
\colhead{R.A.(2000)} & \colhead{decl.(2000)}  \\
}
\startdata

$129{\fdg}1456$	        &  $-13{\fdg}0579$            \\
$129{\fdg}1619$       &  $-13{\fdg}0663$       	 \\
$129{\fdg}0891$         &  $-13{\fdg}0854$		 \\
$129{\fdg}1054$	        &  $-13{\fdg}0938$     	 \\
\enddata

\label{error}
\end{deluxetable}

\clearpage 

\begin{deluxetable}{ccccc}
\tabletypesize{\footnotesize}
\tablecaption{\it Spectral Fits to the Prompt Emission of GRB~001025A$^a$}
\tablewidth{0pt}
\tablehead{
\colhead{Interval} & \colhead{Normalization}                  &\colhead{$\alpha$} &\colhead{$\beta$} &\colhead{$\rm E_0$} \\
\colhead{(s)}         & \colhead{($\rm photons \, cm^{-2}\, s^{-1} $)}   
&\colhead{}            &\colhead{  }         &\colhead{(keV)}           \\
}
\startdata

0--0.256	        &   0.26 $\pm 0.1$             			&    +0.81 $\pm$ 0.57	 & -1.84 $\pm$ 0.12          &	85 $\pm$ 33  \\
0.256--0.768            &   0.37 $\pm 0.04$			&    -0.39 $\pm$ 0.11	 & -2.75 $\pm$ 0.2         &  140 $\pm$ 19 \\
0.768--8.960            &   0.03 $\pm 0.1$					&    -0.99 $\pm$ 0.3	 & -2.40 $\pm$ 0.2         &  103 $\pm$ 46 \\
0--0.768	        &   0.29 $\pm 0.04$ 				&    -0.30 $\pm$ 0.12	 & -2.55 $\pm$ 0.16        &	148 $\pm$ 20 \\
0--8.960                &   0.036 $\pm 0.004$ 				&    -0.91 $\pm$ 0.1     & -2.40 $\pm$ 0.16        &  191 $\pm$ 37 \\ 

\enddata
\tablenotetext{a}{Note: See Eq.~\ref{eqband} for the definition of $\alpha$, $\beta$, and E$_0$.}
\label{konus_spec_tab}
\end{deluxetable}

\clearpage

\begin{deluxetable}{lcccc}
\tabletypesize{\footnotesize}
\tablecaption{\it Journal of VLT Observations}
\tablewidth{0pt}
\tablehead{
\colhead{UT} & \colhead{$\Delta t$} & \colhead{Filter} &  \colhead{Exp. time} & 
\colhead{Seeing} \\
             & \colhead{(days)}   &                  &  \colhead{(s)}       & 
\colhead{(arcsec)} \\
}
\startdata

2000 Oct 26.34 &   1.21 & $R$  & 12$\times$200 & 0.50--0.75 \\
2000 Oct 27.32 &   2.19 & $R$  &  8$\times$200 & 0.50--0.75 \\
2001 Mar 23.02 & 146.89 & $B$  &  3$\times$200 & 0.45--0.55 \\
2001 Mar 23.04 & 146.91 & $R$  & 10$\times$200 & 0.45--0.55 \\

\enddata
\label{vltlog}

\end{deluxetable}

\clearpage 

\begin{deluxetable}{lccc}
\tabletypesize{\footnotesize}
\tablecaption{\it {\it XMM-Newton} Observations of GRB~001025A}
\tablewidth{0pt}
\tablehead{
\colhead{Date\tablenotemark{a}} & \colhead{Time since GRB}  &\colhead{Instrument} &
\colhead{Exposure time\tablenotemark{b}}  \\
\colhead{}     & \colhead{(days)              }  &\colhead{ }          
&\colhead{(ks)}         \\
}
\startdata

2000 Oct 27.003\tablenotemark{c}  & 1.88      & pn           & 15      \\

                                  &           & MOS          & 32      \\

2002 Nov 25.16\tablenotemark{d}   & 761       & pn           & 8.8     \\
        
                                  &           & MOS          & 8.6     \\

2003 Apr 23.29\tablenotemark{e}   & 910       & pn           & 18      \\
 
                                  &           & MOS          & 19      \\
\enddata
\tablenotetext{a}{Start of observation}
\tablenotetext{b}{Exposure time for each detector used in the analysis, 
i.e. with high background periods left out}
\tablenotetext{c}{Target of opportunity observation \citep{alta,altb}}
\tablenotetext{d}{AO-2 observation}
\tablenotetext{e}{Continuation of AO-2 observation}

\label{xmmlog}

\end{deluxetable}

\clearpage

\begin{deluxetable}{lcccc}
\tabletypesize{\footnotesize}
\tablecaption{\it Synchrotron Model Fits to the X-ray Afterglow Spectrum of 
GRB~001025A}
\tablewidth{0pt}
\tablehead{
\colhead{Model} & \colhead{$\Gamma^a$} & \colhead{Absorption$^{b}$} & \colhead{Redshift$^c$} & \colhead{$\chi^2$/d.o.f.}\\
\colhead{}              & \colhead{}      & \colhead{($10^{21}$~cm$^{-2}$)}  & \colhead{} & \colhead{}\\
}
\startdata

Fixed index, fixed Galactic abs.           &  2.0           & 0.61  & 0 & 66.8/33 \\
Free Galactic abs.            &  $2.8 \pm 0.3$ & $2.6^{+0.4}_{-0.6}$ & 0 & 32.6/32 \\
Fixed Galactic abs, free abs. at z       &  $2.6 \pm 0.3$ & $15^{+26}_{-5}$ & $<7.5$ & 31.3/31 \\
\enddata

\tablenotetext{a}{\rm{Photon index.}}
\tablenotetext{b}{\rm{Column density of absorber at the quoted redshift.}}
\tablenotetext{c}{\rm{Redshift of absorber.}}

\label{PL}

\end{deluxetable}

\clearpage 

\begin{deluxetable}{lcccc}
\tabletypesize{\footnotesize}
\tablecaption{\it Thermal Emission Model Fits to the X-ray Afterglow 
Spectrum of GRB~001025A}
\tablewidth{0pt}
\tablehead{
\colhead{Model} & \colhead{T} & \colhead{Absorption$^a$} & 
\colhead{Redshift$^b$} & \colhead{$\chi^2$/d.o.f.}\\
\colhead{}              & \colhead{(keV)}      
& \colhead{($10^{21}$~cm$^{-2}$)}  & \colhead{} & \colhead{}\\
}
\startdata

Fixed Gal. abs.           &  $3.0^{+0.5}_{-0.4}$ & 0.61 & $0.32\pm 0.05$ & 36.7/32\\
Free Gal. abs.            &  $2.8 \pm 0.4$ & $0.9^{+0.4}_{-0.3}$ & $0.33\pm 0.05$ & 
34.1/31 \\
Fixed Gal. abs, free abs. at z  &  $2.8 \pm 0.5$ & $0.5^{+0.7}_{-0.5}$ & $0.33\pm0.05 $ & 
33.7/31 \\

\enddata

\tablenotetext{a}{\rm{Absorber column density.}}
\tablenotetext{b}{\rm{Common redshift of thermal emission and extra-galactic 
absorber.}}

\label{mekal}

\end{deluxetable}

\clearpage

\begin{deluxetable}{lccccc}
\tabletypesize{\footnotesize}
\tablecaption{\it Combined Synchrotron Model and Thermal Emission Model Fits to 
the X-ray Afterglow Spectrum of GRB~001025A}
\tablewidth{0pt}
\tablehead{
\colhead{Model} & \colhead{$\Gamma^a$} & \colhead{T} & \colhead{Absorption$^b$} &  
\colhead{Redshift$^c$} & \colhead{$\chi^2$/d.o.f.}\\
\colhead{}              & \colhead{}              & \colhead{(keV)}      
& \colhead{($10^{21}$~cm$^{-2}$)}  & \colhead{} & \colhead{}\\
}
\startdata

Fixed index, free Gal. abs.  & 2.0 & $1.5\pm 0.6$ & $3.0^{+1.4}_{-1.0}$ & $1.98^{+0.13}_{-0.15}$ & 17.9/27\\

Fixed Gal. abs. & $1.2^{+1.6}_{-1.5}$  & $2.8^{+0.5}_{-0.7}$ & 0.61  & $0.32\pm 0.05$ & 34.1/29\\

Free Gal. abs.  &  $1.9^{+0.6}_{-0.4}$ & $1.6^{+0.5}_{-0.4}$ & $2.8^{+1.0}_{-0.7}$ & $2.00^{+0.12}_{-0.15}$ & 18.4/28\\

Fixed Gal. abs, free abs. at z  &  $1.8 \pm 0.6$ & $1.8\pm 0.6$ & $17^{+9}_{-6}$ & 
$2.00\pm 0.13$ & 19.6/28\\

Fixed index, free abs. at z & 2.0 & $1.7\pm 0.5$ & $17\pm 6$ & $2.00^{+0.12}_{-0.14}$ & 19.8/29\\

\enddata

\tablenotetext{a}{\rm{Photon index.}}
\tablenotetext{b}{\rm{Absorber column density.}}
\tablenotetext{c}{\rm{Common redshift of thermal emission and extra-galactic 
absorber.}}

\label{PLmekal}

\end{deluxetable}

\clearpage


\begin{thebibliography}{}

\bibitem[Altieri et al.(2000a)]{alta}Altieri, B., Schartel, N., Santos, M., 
Tomas, L., Guainazzi, M., Piro, L., \& Parmar, A. 2000a, GCN Circ. 869 
(http://gcn.gsfc.nasa.gov/gcn/gcn3/869.gcn3)

\bibitem[Altieri et al.(2000b)]{altb}Altieri, B., Schartel, N., Lumb, D., 
Piro, L., \& Parmar, A. 2000b, GCN Circ. 884 
(http://gcn.gsfc.nasa.gov/gcn/gcn3/884.gcn3)

\bibitem[Atteia (2003)]{att}Atteia, J.-L. 2003, \aap \, 407, L1

\bibitem[Band et al.(1993)]{band}Band, D., et al. 1993, \apj \, 413, 281

\bibitem[Berger et al.(2002)]{berger}Berger, E., et al. 2002, \apj \, 581, 981 

\bibitem[Berger et al.(2003)]{berger03}Berger, E., Kulkarni, S.R., Frail, D.A., 
2003, \apj \, 590, 379 

\bibitem[Berger et al.(2005)]{berger05}Berger, E., et al. 2005, \apj \, submitted, 
astro-ph/0505107 

\bibitem[Bloom, Frail \& Kulkarni(2003)]{bloom}Bloom, J.S., Frail, D.A. \& 
Kulkarni, S.R. 2003, \apj \, 594, 674 

\bibitem[Bloom et al.(2005)]{bloom05}Bloom, J.S., et al. 2005, \apj \, 
submitted (astro-ph/0505480) 

\bibitem[Castro Cer\'{o}n et al.(2004)]{josemaria}Castro Cer\'{o}n, J.M., et al. 
2004, \aap \, 424, 833

\bibitem[Costa(1999)]{costa}Costa, E., 1999, A\&AS 138, 425

\bibitem[Chincarini et al.(2005)]{chin}Chincarini, G. et al. 2005, astro-ph/0506453

\bibitem[Crew et al.(2003)]{crew}Crew, G. et al. 2003, \apj \, 599, 387

\bibitem[De Pasquale et al.(2003)]{depas}De Pasquale, M. et al., 2003, \apj \, 
592, 1018

\bibitem[Dickey \& Lockman(1990)]{dicklock}Dickey, J.M. \& Lockman, F.J.,
1990, ARA\&A, 28, 215

\bibitem[Djorgovski et al.(2001)]{djorg}Djorgovski, S.G., et al. 2001, \apj \, 
562, 654 

\bibitem[Fenimore et al.(1996)]{fen}Fenimore, E.E., et al. 1996, 
\apj \, 473, 998

\bibitem[Fynbo et al.(2000)]{fynbo}Fynbo, J. P. U., et al. 2000, GCN Circ. 867
         (http://gcn.gsfc.nasa.gov/gcn/gcn3/867.gcn3)

\bibitem[Fynbo et al.(2001)]{fyn01}Fynbo, J. P. U., et al. 2001, 
\aap \, 369, 373

\bibitem[Galama \& Wijers(2001)]{galwij}Galama, T.J. \& Wijers, R.A.M.J. 2001, 
\apj \, 549, L209

\bibitem[Gehrels et al.(2005)]{gehrels}Gehrels, N., et al. 2005, \nat \, 
submitted (astro-ph/0505630)

\bibitem[Giblin et al.(2002)]{giblin}Giblin, T.W., et al. 2002, \apj \, 
570, 573 \aap \, 383, 112

\bibitem[Gorosabel et al.(2002)]{gorosabel02}Gorosabel, J., et al. 2002, 
\aap \, 383, 112

\bibitem[Groot et al.(1998)]{groot98}Groot, P., et al. 1998, 
\apj \, 493, L27 

\bibitem[Hjorth et al.(2002)]{hjorth02}Hjorth, J., et al. 2002, 
\apj \, 576, 113 

\bibitem[Hjorth et al.(2003a)]{hjorth03a}Hjorth, J., et al. 2003a, 
\apj \, 597, 699

\bibitem[Hjorth et al.(2003b)]{hjorth03b}Hjorth, J., et al. 2003b, 
\nat \, 423, 847

\bibitem[Hjorth et al.(2005)]{hjorth05}Hjorth, J., et al. 2005, 
\apj \, submitted (astro-ph/0506123)

\bibitem[Hurley, Cline \& Smith(2000)]{hur00a}Hurley, K., Cline, T., \& 
Smith, D. 2000, GCN Circ. 863 (http://gcn.gsfc.nasa.gov/gcn/gcn3/863.gcn3)

\bibitem[Hurley(2000)]{hur00b}Hurley, K. 2000, GCN Circ. 864,
              (http://gcn.gsfc.nasa.gov/gcn/gcn3/864.gcn3)

\bibitem[Hurley et al.(2002)]{hur02a}Hurley, K. et al. 2002, \apj \, 567, 447

\bibitem[Jakobsson et al.(2004)]{pall}Jakobsson, P. et al. 2004, \apj \, 617, L21

\bibitem[Jakobsson(2005)]{pallhtml}Jakobsson, P. 2005, http://www.astro.ku.dk/\~{}pallja/dark.html

\bibitem[Kumar \& Panaitescu(2000)]{kp}Kumar, P., \& Panaitescu, A. 2000, \apj \, 541, L51
 
\bibitem[Lamb \& Reichart(2000)]{lr}Lamb, D.Q., \& Reichart, D. 2000, \apj \, 536, 1

\bibitem[Lazzati et al.(2002)]{lazatti02}Lazatti, D., Covino, S., \& Ghisellini, G.,
2002, \mnras \, 330, 583

\bibitem[Liedahl, Osterheld \& Goldstein(1995)]{lied}Liedahl, D.A., 
Osterheld, A.L., \& Goldstein, W.H., 1995, \apj \, 438, L115

\bibitem[Malesani et al.(2004)]{malesani}Malesani, D., et al. 2004, 
\apj \, 609, L5

\bibitem[Meszaros(2002)]{mesz}Meszaros, P. 2002, ARAA 40, 137

\bibitem[Mewe et al.(1985)]{mewe}Mewe, R., Gronenschield, E.H.B.M., \&
van den Oord, G.H.J., 1985, A\& AS \, 62, 197

\bibitem[Piro(2001)]{piro01}Piro, L., 2001, in Gamma-ray bursts in the 
afterglow era, eds. E Costa,
F. Frontera, \& J. Hjorth (Springer - Berlin), p. 97

\bibitem[Piro et al.(2002)]{piro}Piro, L., et al., 2002, \apj \, 577, 680

\bibitem[Predehl \& Schmitt(1995)]{ps}Predehl, P. \& Schmitt, J.H.M.M., 1995, 
\aap \, 293, 889

\bibitem[Preece et al.(1998)]{preece}Preece, R., Pendleton, G., Briggs, M., 
Mallozzi, R., Paciesas, 
W., Band, D., Matteson, J., \& Meegan, C. 1998, \apj \, 496, 849

\bibitem[Ramirez-Ruiz et al.(2002)]{rr}Ramirez-Ruiz, E., et al., 2002, \mnras \, 337, 1349

\bibitem[Reichart \& Price(2002)]{rp}Reichart, D., \& Price, P. 2002, \apj \, 565, 174

\bibitem[Rol et al.(2005)]{rol}Rol, E. et al. 2005, accepted for publication in \apj

\bibitem[Reeves et al.(2002)]{reeves}Reeves, J. et al. 2002, \nat \, 416, 512

\bibitem[Sari, Piran \& Narayan(1998)]{spn}Sari R., Piran, T. \& Narayan, R., 1998, 
\apjl \, 497, L17

\bibitem[Schlegel et al.(1998)]{schegel}Schlegel, D.J. et al. 1998, 
\apj \, 500, 525

\bibitem[Smith et al.(2000)]{smith}Smith, D.A., Levine, A., Remillard, R., Hurley, 
K., \& Cline, T. 2000,
	GCN Circ. 861 (http://gcn.gsfc.nasa.gov/gcn/gcn3/861.gcn3)

\bibitem[Smith et al.(2002)]{smith02}Smith D.A., et al., 2002,
\apjs \, 141, 415

\bibitem[Stanek et al.(2003)]{stanek}Stanek, K.Z., et al., 2003,
\apj \, 591, L17

\bibitem[Tagliaferri et al.(2003)]{tag}Tagliaferri, G., et al., \nat \, accepted,
astro-ph/0506355

\bibitem[Taylor et al.(1998)]{taylor98}Taylor, G.B., et al., 1998,
\apj \, 502, L115

\bibitem[Uemura et al.(2000)]{uemura}Uemura, M., Kato, T., Ishioka, R., Iwamatsu, H., 
\& Yamaoka, H. 2000
        GCN Circ. 866 (http://gcn.gsfc.nasa.gov/gcn/gcn3/866.gcn3)

\bibitem[Watson et al.(2002)]{watson}Watson, D., Reeves, J., Osborne, J., O'Brien, P., 
Pounds, K., Tedds, J.,          
          Santos-Lleo, M., \& Ehle, M. 2002 \aap \, 393, L1

\bibitem[Watson et al.(2003)]{watson03}Watson, D., Reeves, J., Hjorth, J., 
\& Pedersen, K. 2003 \apjl \, 595, L29

\bibitem[Zhang \& Meszaros(2004)]{zhangmes}Zhang, B. \& Meszaros, P. 2004 
Int. J. Mod. Phys. A \, 19, 2385

\end{thebibliography}
\end{document}